\documentclass[twocolumn,superscriptaddress,showpacs,prx]{revtex4-2}
\usepackage[latin9]{inputenc}
\pdfoutput=1
\usepackage[english]{babel}
\usepackage[T1]{fontenc}
\usepackage{hyperref}
\usepackage{orcidlink}

\usepackage{wrapfig}

\usepackage{amsmath, amssymb, amstext, amsfonts}
\usepackage{amsthm}
\usepackage{float}
\usepackage{graphicx}
\usepackage{physics}
\usepackage{xcolor}
\usepackage[shortlabels]{enumitem}

\def\R{\mathbb{R}}

\def\be{\begin{equation}}
\def\ee{\end{equation}}
\def\bea{\begin{eqnarray}}
\def\eea{\end{eqnarray}}
\def\bma{\begin{mathletters}}
\def\ema{\end{mathletters}}

\def\q0{\underline{0}}

\def\C{{\mathbb C}}

\def\R{\mathbb{R}}

\newtheorem{observation}{Observation}

\def\one{\leavevmode\hbox{\small1\normalsize\kern-.33em1}}

\def\one{\leavevmode\hbox{\small1\normalsize\kern-.33em1}}

\newcommand{\ba}{\begin{eqnarray}}
\newcommand{\ea}{\end{eqnarray}}
\newcommand{\ban}{\begin{eqnarray*}}
\newcommand{\ean}{\end{eqnarray*}}

\begin{document}


\title{Beating one bit of communication with and without quantum pseudo-telepathy}
\author{Istv\'an M\'arton}
\affiliation{MTA Atomki Lend\"ulet Quantum Correlations Research Group, Institute for Nuclear Research, P.O. Box 51, H-4001 Debrecen, Hungary} 
\author{Erika Bene}
\affiliation{MTA Atomki Lend\"ulet Quantum Correlations Research Group, Institute for Nuclear Research, P.O. Box 51, H-4001 Debrecen, Hungary}
\author{P\'eter Divi\'anszky}
\affiliation{MTA Atomki Lend\"ulet Quantum Correlations Research Group, Institute for Nuclear Research, P.O. Box 51, H-4001 Debrecen, Hungary}
\author{Tam\'as V\'ertesi}
\affiliation{MTA Atomki Lend\"ulet Quantum Correlations Research Group, Institute for Nuclear Research, P.O. Box 51, H-4001 Debrecen, Hungary} 

\begin{abstract}
According to Bell's theorem, certain entangled states cannot be simulated classically using local hidden variables (LHV). But if can we augment LHV by classical communication, how many bits are needed to simulate them? There is a strong evidence that a single bit of communication is powerful enough to simulate projective measurements on any two-qubit entangled state.
In this study, we present Bell-like scenarios where bipartite correlations resulting from projective measurements on higher dimensional states cannot be simulated with a single bit of communication. These include a three-input, a four-input, a seven-input, and a 63-input bipartite Bell-like inequality with 80089, 64, 16, and 2 outputs, respectively.
Two copies of emblematic Bell expressions, such as the Magic square pseudo-telepathy game, prove to be particularly powerful, requiring a $16\times 16$ state to beat the one-bit classical bound, and look a promising candidate for implementation on an optical platform. 
\end{abstract}

\maketitle

\section{Introduction}

Certain mutipartite quantum correlations cannot be simulated by local hidden variables (LHV), also known as shared random variables. This forms the core of Bell's theorem~\cite{Bell1964,CHSH1969}. When a quantum correlation cannot be simulated by LHV models, it is referred to Bell nonlocal~\cite{Brunner2014}. One question that arises is: Which resources are needed on top of LHVs to simulate quantum correlations? The most obvious resource is LHV, augmented by classical communication~\cite{Maudlin1992,Brassard1999,Cerf2000,Steiner2000,Csirik2002}. In particular, the question can be turned into a quantitative one: at least how many bits of classical communication are required to reproduce Bell nonlocal correlations arising from any number of measurements on $d\times d$ quantum states. It is worth noting that Bell nonlocal correlations have also found a crucial role in applied physics, e.g. they can be used for the device-independent certification of the correct functioning of quantum key distribution~\cite{Acin2007QKD}, random number generators~\cite{Pironio2010QRNG} and other devices (see e.g. Ref~\cite{Scarani2019} for a thorough review of the field). However, there exist less strict frameworks that include partially characterized devices and the classical communication costs of these protocols have also been studied (see, for example, references for bipartite systems~\cite{Nagy2016,Sainz2016,Brask2017} and for single systems \cite{Cerf2000,Renner2023}).  

First, the communication cost of simulating maximally entangled states was addressed. Following initial results~\cite{Brassard1999,Csirik2002}, it has been proven that projective measurements on a two-qubit maximally entangled state can be simulated with LHV augmented by one bit of classical communication (let us call it the one-bit classical model)~\cite{Toner2003}. What if the qubits are partially entangled? Research has been shown that projective measurements on all partially entangled two-qubit states can be classically simulated  with at most two bits of communication~\cite{Toner2003}. However, Gisin has posed the question whether one bit is sufficient~\cite{Gisin2007open}. 
This problem can be systematically approached, as the one-bit classical resources are contained within a Bell-type polytope~\cite{Bacon2003,Maxwell2014,Renner2023}. However, the size of the one-bit classical polytope grows rapidly with the number of inputs and outputs. Currently, the largest completely characterized one-bit classical polytope has three measurements for one party, two measurements for the other party, and binary outputs~\cite{Maxwell2014}. No quantum violation has been found, even for three inputs per party on both sides~\cite{ZambriniCruzeiro2019}. Recently, the problem was approached from a different viewpoint and aimed to simulate two-qubit states with an arbitrary number of projective measurements. In particular, Renner and Quintino~\cite{Renner2022} devised a one-bit classical protocol that perfectly simulates projective measurements on weakly entangled two-qubit states. Sidayaja et~al.'s~\cite{Sidajaya2023} recent numerical study using neural networks has gathered strong evidence that projective measurements on all two-qubit states can be simulated with a one-bit classical model.

The strategy in this study is to identify Bell-like inequalities that are satisfied by all LHV models supplemented with one bit of communication (i.e., one-bit classical models), and then search for a quantum violation of these inequalities. As the two-qubit scenario has been widely studied without violation of the one-bit classical model, here we turn to higher-dimensional bipartite systems. What are the perspectives of solving this problem? On one hand, complexity arguments show that one bit of communication is not sufficient to simulate all bipartite quantum correlations classically~\cite{Brassard1999}. On the other hand, it is an open problem to identify such Bell-type scenarios with a modest number of inputs and outputs (see e.g. Sidajaya et al.~\cite{Sidajaya2023}) For instance, it is known that correlations resulting from two-output measurements on arbitrary high-dimensional maximally entangled states can be simulated classically by using a mere two bits of communication~\cite{Regev2010}. In fact, this bound is tight, since there exist $4\times 4$ dimensional quantum states that cannot be simulated with a single bit of communication. However, the proof involves an infinite number of measurement inputs~\cite{Vertesi2009}.

\begin{table*}[t]
\begin{center}
\begin{tabular}{ |c|r|c|c|r|r|r|r| } 
 \hline
 Design & Section & ($m_A,m_B,o_A,o_B$) & Direction & $L1\text{bit}$ & $Q$ & $d$ & $D_P$\\
 \hline\hline
 $\text{CHSH}^{\otimes4}$ & \ref{sec:CHSH1bit} & $(16,16,16,16)$ & bi & $132^*$ &  135.8822 & 16 & 65536\\ 
 $\text{Magic}^{\otimes2}$ & \ref{sec:magic1bit} & $(9,9,16,16)$ & bi & 75\,\; & 81 & 16 & 20736 \\ 
 $[\text{Magic}^{\otimes 2}]_s$ & \ref{sec:magic1bittrunc}  & $(7, 7, 16, 16)$ & bi & 48\,\; & 49 & 16 & 12544\\
$\text{CGLMP}_8^{\otimes2}$ & \ref{sec:cglmp1bit}    &    $(4, 4, 64, 64)$   &   bi  &  12\,\;   &  12.1230     &  64   &   65536\\
$[\text{CGLMP}_{283}^{\otimes2}]_s$ & \ref{sec:cglmp1bittrunc} & $(3, 3, 283^2, 283^2)$   &     bi     &      8\,\;  &  8.0002059   &  80089   &   $240267^2$ \\
$[\text{Magic}^{\otimes2}]_a$ & \ref{sec:magic1bittruncasym}   &   $(7, 3, 16, 16)$    &  fixed  & 20\,\;  &   21  &  16   &     5376\\
$[\text{CGLMP}_{38}^{\otimes2}]_a$ & \ref{sec:cglmp1bittruncfixed} &  $(3, 2, 38^2, 38^2)$    &  fixed  &   5\,\; &   5.0005456  &   1444  &  12510816\\
$\text{Plato}_{E7}$ & \ref{sec:platonic} &   $(63, 63, 2, 2)$   & bi  &   $<565$\,\;  &  567  &         2   &  15876\\
\hline\hline
\end{tabular}
\end{center}
\caption{\label{tab_I} The term ``Design'' refers to the construction, denoted together with a section number. The scenario is denoted by ($m_A,m_B,o_A,o_B$). ``Direction'' refers to the direction of the one-bit communication and can either be bi-directional or a fixed one-directional. In addition, the table shows for each construction the one-bit classical bound ($L\text{1bit}$), the quantum value $Q$, the dimension $d$ of the component space and the dimension $D_P$ of the full probability space. The entry marked by $(^*)$ indicates that the one-bit bound comes from an extensive, though not rigorous, computation. All other entries are exact.}
\end{table*}

This paper presents several examples that beat quantumly the one-bit classical bound with a finite and typically a modest number of measurement inputs and outputs. Our tools are based on four bipartite Bell inequalities: the CHSH inequality~\cite{CHSH1969}, the Magic square game~\cite{Mermin1990,Cabello2001,Aravind2002}, the family of CGLMP inequalities~\cite{CGLMP2002}, and Platonic Bell inequalities~\cite{Tavakoli2020,BolonekLason2021,Pal2022}. We use them as building blocks to our Bell-type constructions.

Table~\ref{tab_I} collects the Bell-type constructions of this paper. We present the setup involving the input cardinality ($m_A$ and $m_B$), output cardinality ($o_A$ and $o_B$), and the $d\times d$ quantum state. Additionally, we provide the one-bit classical bound ($L1bit$), the quantum value ($Q$) of the Bell-type inequalities, and $D_P = m_Am_Bo_Ao_B$. We use this as a measure of complexity of the given Bell-type construction. 
Note that $D_P>24$ represents a lower bound to surpass the one-bit classical bound quantumly. This is due to the fact that the non-trivial probability distributions that have at most dimension $D_P=24$ are given by the scenarios $(2,3,2,2)$, $(3,2,2,2)$, $(2,2,3,2)$, and $(2,2,2,3)$, and all of them can be simulated classically with one bit of bidirectional communication (i.e. communication either from Alice to Bob or vice versa).

It should be noted that the $L1\text{bit}$ results presented in Table~\ref{tab_I} are the result of rigorous computation except for the case of $\text{CHSH}^{\otimes 4}$, for which the $L1\text{bit}$ bound is obtained from heuristics. As we see, all $D_P$ values are much greater than 24, demonstrating the power of a single bit of classical communication, or alternatively, our lack of success in finding constructions that exceed the one-bit bound with lower complexity. We propose it as an open problem to shrink the gap between 24 and 12544.

Table~\ref{tab_I} reveals that the smallest $D_P$ is given by the $[\text{Magic}^{\otimes 2}]_s$ inequality. The inequality features seven inputs and sixteen outputs. We propose it as a candidate to experimentally violate the one-bit classical bound. If we aim to violate quantumly the fixed-directional (e.g., from Alice to Bob) one-bit bound, then the most suitable candidate appears to be the $[\text{Magic}^{\otimes 2}]_a$ inequality. The inequality comprises seven inputs on Alice's side and three inputs on Bob's side, with sixteen outputs per measurement on each side. 

Furthermore, we provide Bell-like inequalities augmented by $c$ bits of one-way classical communication. These examples are based on multiple copies of the CHSH expression (as discussed in section~\ref{sec:CHSHcbit}), as well as a truncated version of multicopy $\text{CGLMP}_d$ inequalities (as discussed in section~\ref{sec:cglmp1bittrunccbit}). We examine the scaling of the input and output cardinality and the Hilbert space dimension to beat the one-way $c$-bit classical bound of the aforementioned inequalities. Crucially, we find a Bell inequality with $c$ bits of one-way classical communication that has $2^c+1$ inputs and can be violated with high-dimensional quantum systems. However, we were unable to provide an explicit lower bound on the dimension required to exceed the $c$-bit bound. Note that this is a minimal scenario with respect to the number of inputs, otherwise the inequality cannot be violated.     

The paper is structured as follows. Section~\ref{sec:CHSH} provides notation, defines the Bell-like scenario augmented by one bit and $c$ bits of communication, and demonstrates the usefulness of multiple copies of the quantum CHSH game in this problem. Section~\ref{sec:magic} considers the double Magic square game and its different truncated versions. It also provides the one-bit classical bound of the associated Bell-type inequalities. In particular, this section shows a quantum violation of the one-bit classical bound of a 7-setting and 16-outcome Bell-type inequality. Section~\ref{sec:cglmp} examines the $\text{CGLMP}_d$ inequalities and their truncated versions, and we demonstrate quantum violation of a three-input $283^2$-output Bell-like inequality with one bit of communication. Section~\ref{sec:platonic} investigates a different construction based on the so-called Platonic Bell inequalities that belong to a class of correlation-type Bell inequalities. We compute the one-bit classical bound for such Bell-type inequalities with 63 inputs numerically and provide an analytical upper bound as well. We then find genuine quantum violation of this bound with $8\times 8$ dimensional quantum states. Section~\ref{sec:disc} discusses the results obtained and future research directions.

\section{The power of multiple copies of the CHSH expression}
\label{sec:CHSH}

\subsection{Notation and the one-bit Bell-type scenario}
\label{sec:CHSHnotation}

In this subsection we establish notation and introduce the concept of Bell-like inequalities that are valid for all correlations that can be simulated classically with a single bit of communication. Consider a scenario where two non-communicating parties, named Alice and Bob, produce outcomes (alternatively outputs) $a\in\{0,\ldots,o_A-1\}$ and $b\in\{0,\ldots,o_B-1\}$ for settings (alternatively inputs) $x\in\{0,\ldots,m_A-1\}$ and $y\in\{0,\ldots,m_B-1\}$. In such a scenario, a generic bipartite Bell inequality can be expressed as
\begin{equation}
{\cal B} = \sum_{a,b,x,y} S_{abxy}P(ab|xy) \le L, 
\label{BI}
\end{equation}
where $P(ab|xy)$ represents the conditional probabilities and we assume that $S_{abxy} \ge 0$. 
Writing a Bell inequality in this form, one can also view it as a bipartite Bell nonlocal game~\cite{Cleve2004}. What is the local bound $L$, which appears on the right-hand side of Eq.~(\ref{BI})? It is the maximal value of the Bell expression $\cal B$ when probabilities $P(ab|xy)$ admit an LHV model. In this case, $P(ab|xy)$ can be explained using a common past history and local operations by Alice and Bob, and can be written as follows:
\begin{equation}
P(ab|xy) = \int q(\lambda)P_A(a|x\lambda)P_B(b|y\lambda).    
\label{P_LHV}
\end{equation}
Here $\lambda$ represents a local variable, $q(\lambda)$ denotes a probability distribution, and $P_A$ and $P_B$ refer to Alice's and Bob's respective marginals. 

Let us now allow one bit of classical communication, say from Alice to Bob, in addition to LHV operations. In that case, the protocol follows these steps. First, Alice and Bob receive their inputs $x$ and $y$. Then, Alice is allowed to send one bit of classical communication, $l=0,1$, to Bob. Afterward, Alice and Bob produce the respective outputs $a$ and $b$. In this way, Alice and Bob can simulate all $P(ab|xy)$ that satisfy:
\begin{equation}
P(ab|xy) = \int q(\lambda)P_A(a|x\lambda)P_B(b|yl\lambda),
\label{P_LHV1bit}
\end{equation}
where the marginal of Bob ($P_B$) also depends on the value of the classical bit $l = l(x,\lambda)$, $l$ being either 0 or 1. The maximum on $\cal B$ in Eq.~(\ref{BI}) achieved by these strategies is referred to as $L1\text{bit}$. 
  
When the probabilities are obtained from quantum mechanics instead, the maximum value of $\cal B$ in Eq.~(\ref{BI}) is called the Tsirelson bound~\cite{Cirelson1980}. In such a case, 
\begin{equation}
P(ab|xy) = \tr(\rho A_{a|x}\otimes B_{b|y}),    
\label{P_Quantum}
\end{equation}
where $\rho$ is a density matrix on the space $\C^d \otimes \C^d$, and $A$ and $B$ are $d$-dimensional projective matrices. These matrices add up to the identity, $\sum_a A_{a|x} = \sum_b B_{b|y} = \one_d$. 

Let us write the CHSH inequality~\cite{CHSH1969} in the following form (see e.g.~\cite{CGLMP2002,Barrett2002}):
\begin{align}
\text{CHSH} =& P(a = b|00) + P(a = b|01)\nonumber \\
&  + P(a = b|10) + P(a \neq b|11) \le 3,
\label{CHSHineq}
\end{align}
where $x,y$ and $a,b$ are assumed to have values of 0 and 1. On the right-hand side of Eq.~(\ref{CHSHineq}), $L=3$ is the local bound, which can be attained by suitable local deterministic strategies. An appropriate strategy is for Alice to output $a=1$ for $x=0,1$, while Bob outputs $b=1$ for $y=0,1$. Thus, the correlations within the local set~(\ref{P_LHV}) can be expressed as $P_L(ab|xy) = \delta_{a,1}\delta_{b,1}$ for any $x,y$.

On the other hand, $L1\text{bit}(\text{CHSH}) = 4$, which can be achieved when Alice sends $l = x$ to Bob, with Alice outputting $a=1$ for $x=0,1$ and Bob outputting $b=1$ for $y=0$ and $b=1-l$ for $y=1$. It is worth noting that 4 is also the algebraic bound that can be achieved with $P(ab|xy)$, solely respecting positivity.

In the quantum case, using a two-qubit maximally entangled state and mutually unbiased measurements, the following statistics can be obtained by~(\ref{P_Quantum}): 
\begin{equation}
P_Q(a,b|x,y) = \frac{1+(\sqrt{2}/2)(-1)^{a\oplus b}(-1)^{xy}}{4}.
\end{equation}
By substituting these values into the CHSH inequality~(\ref{CHSHineq}), one obtains $Q(\text{CHSH}) = 2 + \sqrt{2}$. 

When given $n$ instances of a Bell nonlocal game $\cal B$, a straightforward way is to play them in parallel. For example, when presented with two copies ($n=2$) and ${\cal B} = \text{CHSH}$, the resulting double CHSH expression~\cite{Barrett2002} (see also Refs.~\cite{Wu2016,Marton2023}) is:
\begin{equation}
\text{CHSH}^{\otimes 2} = \text{CHSH}_{A,B} \otimes \text{CHSH}_{A',B'},   \end{equation}
where $\text{CHSH}_{A,B}$ acts on the first copy, while $\text{CHSH}_{A',B'}$ acts on the second copy of the space of input-output variables. The formula for $n\ge 2$ copies is as follows
\begin{equation}
\text{CHSH}^{\otimes n} = \bigotimes_{i=1}^n \text{CHSH}_{A^{(i)},B^{(i)}}.    
\end{equation}

While computing the Tsirelson bound of ${\cal B}^{\otimes n}$ for a generic $\cal B$ is difficult, we can often obtain a good enough lower bound by playing each instance of $\cal B$ independently with the optimal quantum strategy for the single copy case. In the case of $n$ copies, we then have
\begin{equation}
P(ab|xy)=\prod_{i=1}^n P(a_ib_i|x_iy_i).
\label{Pncopy}
\end{equation}
In general, for the quantum maximum of the $n$-copy Bell functional $\cal B$ we only obtain a lower bound: $Q({\cal B}^{\otimes n}) \ge Q({\cal B})^n$. However, in the particular case of $n$-copy CHSH, the Tsirelson bound saturates the lower bound~\cite{Cleve2008}, and we have
\begin{equation}
Q(\text{CHSH}^{\otimes n}) = (2 + \sqrt{2})^n.
\label{QncopyCHSH}
\end{equation}
What is the local bound of $\text{CHSH}^{\otimes n}$? One obvious lower bound is $L\le 3^n$, which can be achieved with independent classical deterministic strategies between the copies. However, exploiting joint strategies enables better performance. In such a case, Alice's output $a_i$ depends not only on input $x_i$, but also on input $x_{i'}$, where $i'\neq i$. For two and three copies, we respectively have the bounds $L(\text{CHSH}^{\otimes 2})=10$ and $L(\text{CHSH}^{\otimes 3})=31$, which were obtained independently by S. Aaronson and B. Toner. However, only empirical values can be found in the literature for $n>3$~\cite{Araujo2020}. On the other hand, the following upper bound was found in 2014 by A. Ambainis~\cite{Yuen}, building upon Ref.~\cite{Dinur2014}:
\begin{equation}
L(\text{CHSH}^{\otimes n}) \le (1 + \sqrt{5})^n,
\label{L_ambainis}
\end{equation}
which holds for any $n\ge 1$.

\subsection{One-bit classical bound for the multi-copy CHSH scenario} 
\label{sec:CHSH1bit}

We show analytically that a certain number of $n$ copies exist for which quantum correlations exceed the one-bit classical bound, $L1\text{bit}$. Two ingredients are required for the proof. The first one corresponds to the bound in Eq.~(\ref{L_ambainis}). The second one is the relation 
\begin{equation}
L1\text{bit}({\cal B}) \le 2L({\cal B}),    
\label{L1bit_comp_L}
\end{equation}
which holds true for any bipartite Bell expression $\cal B$. By  choosing ${\cal B} = \text{CHSH}^{\otimes 2}$ and combining the two relations~(\ref{L_ambainis},\ref{L1bit_comp_L}), the following upper bound is reached:
\begin{equation}
L1\text{bit}(\text{CHSH}^{\otimes n}) \le 2(1 + \sqrt{5})^n.
\label{L1bit_ncopyCHSH}
\end{equation}

On the other hand, the Tsirelson bound of the $n$-copy CHSH expression is given by Eq.~(\ref{QncopyCHSH}). Applying Eq.~(\ref{L1bit_ncopyCHSH}) results in the value $n = 13$ at which the quantum value $Q$ exceeds the one-bit bound of $L1\text{bit}$. This calculation relied on applying analytical upper bounds. However, is the value $n=13$ tight?  We used heuristic search to compute $L1\text{bit}(\text{CHSH}^{\otimes n})$ for small $n$ and we found that $n = 4$ is the critical value at which $L1\text{bit}(\text{CHSH}^{\otimes n}) < Q(\text{CHSH}^{\otimes n})$. In this Bell-type scenario, each party has $m = 2^n = 16$ measurement inputs, $o = 16$ measurement outputs, and ($16\times 16$)-dimensional states. See Table~\ref{tab_II}. The entries without stars in the table are obtained through exact enumeration, while those marked with stars ($^*$) are based on heuristics. The heuristics for the one-bit classical bound is a modified version of the see-saw procedure, where iteration is also performed for the optimal strategies for the $l(x)=0,1$ bit message, as used e.g. in Refs.~\cite{Divianszky2017,Araujo2020}. Note that, since the $\text{CHSH}^{\otimes n}$ expression is symmetric for party exchange, all the above findings apply to the bidirectional case.

\begin{table}[t]
\begin{center}
\begin{tabular}{ |r|r|r|r| } 
 \hline
$n$ & $L(\text{CHSH}^{\otimes n})$ & $L1\text{bit}(\text{CHSH}^n)$ & $Q(\text{CHSH}^{\otimes n})$\\
 \hline\hline
 1 & 3\,\; & 4\,\; & 3.41421 \\ 
 2 & 10\,\; & 16\,\; & 11.65685 \\ 
 3 & 31\,\; & 40\,\; & 39.79898 \\ 
 4 & $100^*$ & $132^*$ & 135.88225\\
 5 & $310^*$ & $408^*$ & 463.93102\\
 6 & $1000^*$ & $1332^*$ & 1583.95959\\
 \hline\hline
\end{tabular}
\end{center}
\caption{\label{tab_II} The table displays the local bound $L$, the one-bit bound $L1\text{bit}$, and the quantum value $Q$ of the $n$-copy CHSH inequality up to $n=6$. The computation of the one-bit bounds is due to this work. Ara\'ujo et al.~\cite{Araujo2020} computed the local bound for $n>3$. All the entries marked with ($^*$) are based on heuristics. The quantum value $Q$ is given analytically by the formula $(1+\sqrt{2})^n$ shown in Eq.~(\ref{QncopyCHSH}).}
\end{table}

\subsection{One-way $c$-bit classical bound for the $n$-copy CHSH scenario} 
\label{sec:CHSHcbit}

We generalize the one-bit classical result from Sec.~\ref{sec:CHSH1bit} to the exchange of $l = (2^c)$-level classical messages, where $c$ is the number of bits. It should be noted that we are considering only a fixed amount of one-way classical communication between the two parties, but our findings can be extended to a fixed amount of two-way communication as well. The set of possible classical protocols that use at most $c$ bits of communication is the subject of the field of communication complexity~\cite{Kushilevitz1997}. We inquire about the number of $n$ copies of CHSH expressions required to surpass the one-way $c$-bit classical bound with the quantum value~(\ref{QncopyCHSH}). In particular, we give an upper bound on the setup parameters, including the number of inputs $m$, outputs $o$, and the dimension $d$ per party needed to beat the $c$-bit classical bound. 

Let us extend the results obtained in section~\ref{sec:CHSH1bit} on one bit to $c$ bits. Firstly
\begin{equation}
Lc\text{bit}({\cal B}) \le 2^c L({\cal B})     
\label{Lcbit_comp_L}
\end{equation}
and specifically for ${\cal B} = \text{CHSH}^{\otimes n}$ and having applied the upper bound~(\ref{L_ambainis}) to a given $c$ and $n$, we obtain
\begin{equation}
Lc\text{bit}(\text{CHSH}^{\otimes n}) \le 2^c(1 + \sqrt{5})^n.     
\end{equation}
Furthermore, given our condition to exceed the $Lc\text{bit}$ bound:
\begin{equation}
2^c(1 + \sqrt{5})^n < (2+\sqrt{2})^n    
\end{equation}
we arrive at the upper bound for the $n_{\text{crit}}(c)$ value: 
\begin{equation}
 n_{\text{crit}}(c) \le 13\times2^c. 
\end{equation} 
Since for the $n$-copy CHSH expression, the number of inputs, outputs, and dimensionality of the component space is the same (i.e., $2^n$), we obtain the following upper bounds:
\begin{equation}
m_{\text{crit}} = o_{\text{crit}} = d_{\text{crit}} \le 2^{13\times2^c}. 
\label{bounds_mod}   
\end{equation}
Note that a lower bound of $m_{\text{crit}}(c) > 2^c$ follows from Alice simply communicating her own input to Bob using an $l = (2^c)$-level classical message.

Regarding the one-bit classical bound, we ask about the possibility of finding more economical Bell-like inequalities augmented by one bit of communication that can be violated quantumly with fewer inputs, outputs and dimensions. In the next sections, we will improve on the $n$-copy CHSH inequality in all the aforementioned bounds~(\ref{bounds_mod}). Nonetheless, we cannot improve of all three parameters simultaneously. Still is it possible to find tighter upper bounds for $m_{\text{crit}}(c)$, $o_{\text{crit}}(c)$ or $d_{\text{crit}}(c)$ for $c>1$ by using other Bell-like inequalities? We shall present an optimal solution to $m_{\text{crit}}(c)$ based on a truncated version of the $2^c$-copies of the $\text{CGLMP}_d$ inequality. A tight lower bound of $m_{\text{crit}}(c) > 2^c$ can be achieved by using a (possibly huge) $D \times D$ quantum state, where $D = {{d}^2}^c$ and $d$ is large enough (see section~\ref{sec:cglmp1bittrunccbit} for the details). 

In the following sections, we will examine the double Magic square game (Sec.~\ref{sec:magic}), the double $\text{CGLMP}_d$ inequalities (Sec.~\ref{sec:cglmp}), and a Platonic Bell-type inequality ~(Sec.~\ref{sec:platonic}). All of these examples are shown to beat the one-bit classical bound quantumly.

\section{Two copies of the Magic square and pseudo-telepathy games}
\label{sec:magic}

\subsection{One-bit bound on the double Magic square game}
\label{sec:magic1bit}

First let us give a brief description of the single-copy Magic square game~\cite{Mermin1990,Cabello2001,Aravind2002}. This is a nonlocal game for which the Tsirelson bound achieves the algebraic maximum. This game is within the class of quantum pseudo-telepathy games~\cite{Brassard2005}. Each party has three inputs and four outputs. The Bell functional ``Magic'' with the local bound of 8 is written as follows:
\begin{equation}
\text{Magic} = \sum_{x,y=0,1,2}P(a_y=b_x|xy) \le 8,  
\end{equation}
where the parties produce three bits each, which are represented as $a=(a_0a_1a_2)$ and $b=(b_0b_1b_2)$. Additionally, it is assumed that the following conditions hold true: $a_0\oplus a_1\oplus a_2 = 0\, \text{mod}\, 2$ and $b_0\oplus b_1\oplus b_2 = 1\, \text{mod}\, 2$. Due to this parity constraint, the third output becomes unnecessary and each party only requires four outputs: $a=a_0a_1\in\{00,01,10,11\}$ and similarly $b=b_0b_1\in\{00,01,10,11\}$. In the game terminology, the existence of a winning quantum strategy means that quantum physics violates this inequality up to the maximal algebraic value of 9. This violation can be obtained with a $4\times 4$ dimensional maximally entangled state.

\begin{table}[t]
\begin{center}
\begin{tabular}{ |r|l|l|l| } 
 \hline
$n$ & $L(\text{Magic}^{\otimes n})$ & $L1\text{bit}(\text{Magic}^n)$ & $Q(\text{Magic}^{\otimes n})$\\
 \hline\hline
 1 & 8 ~\cite{Aravind2002} & 9 ~\cite{Broadbent2006} & 9 ~\cite{Aravind2002}\\ 
 2 & 66 ~\cite{Araujo2020} & 75 & 81 \\ 
 3 & $528^*$ ~\cite{Araujo2020} & $621^*$ & 729 \\ 
 \hline\hline
\end{tabular}
\end{center}
\caption{\label{tab_III} The table lists the local bound $L$, the one-bit bound $L1\text{bit}$, and the quantum value $Q$ for the $n$-copy Magic square Bell-type inequality for values up to $n=3$. The computation of the local bound is noted in the references, whereas the computation of the one-bit bound for $n=2,3$ is due to the present work. Entries for $L$ and $L1\text{bit}$ without stars are calculated by exact enumeration, whereas entries with a star $(^*)$ are based on heuristics. The quantum value $Q=9^n$ defines a lower bound to the Tsirelson bound for any value of $n$.}
\end{table}

It is known that $L(\text{Magic}) = 8$, $L1\text{bit}(\text{Magic}) = Q(\text{Magic}) = 9$ (see Ref.~\cite{Broadbent2006}). Therefore, there is no advantage of using quantum strategies over the optimal one-bit classical protocol. 
Table~\ref{tab_III} summarizes the different bounds for the $n$-qubit Magic square game up to $n=3$. As we can see, two copies are sufficient to beat the one-bit bound ($L1\text{bit}(\text{Magic}^{\otimes 2}=75)$) quantumly ($Q(\text{Magic}^{\otimes 2}=81)$). The value of 75 has been verified using the branch-and-bound algorithm~(see Ref.~\cite{Divianszky2017}) adapted to the one-bit problem. We implemented this algorithm, which gives the exact one-bit bound, using CPU parallel computation, and reproduced the value of 75 in 11 seconds on our workstation. However, this value can also be proved using analytical arguments. Below, we only prove that $L1\text{bit}(\text{Magic}^{\otimes2})\le 80$, which is strictly less than the algebraic bound of 81. 

\begin{proof}
Denote the set of nine inputs on Alice's side as $X(x,x') = \{0,1,2\}^2$, and use the same set of inputs on Bob's side $Y(y,y') = \{0,1,2\}^2$. In order to prove that the one-bit classical bound for $\text{Magic}^{\otimes2}$ is less than the algebraic bound, we make use of the definition~(\ref{P_LHV1bit}) for the one-bit classical set. We have to consider all possible bipartitions of $X$ according to the classical message $l = 0,1$, add up the local bound for each partition, and choose the partition with the greatest sum to obtain the $L1\text{bit}$ bound. Since we are only concerned whether $L1\text{bit}$ can attain the algebraic maximum or not, it is enough to show that the local bound of one of the partitions cannot attain the algebraic maximum that corresponds to that particular partition. In this context, it is useful to introduce a coarse-graining of the joint probability distribution $P(aa'bb'|xx'yy')$ for a given $x'\in\{0,1,2\}$ on Alice's side and $y'\in\{0,1,2\}$ on Bob's side:
\begin{equation}
P(ab|xy)=\sum_{a',b'}P(aa'bb'|xx'yy'), 
\label{Pabxycg}
\end{equation}
where summation is over all $a'=\{0,1\}^2$ and $b'=\{0,1\}^2$ outputs. Let us observe that a probability distribution $P(ab|xy)$ in Eq.~(\ref{Pabxycg}) that corresponds to the algebraic maximum of a single-copy pseudo-telepathy game does not allow the original two-copy distribution $P(aa'bb'|xx'yy')$ to be achieved via a local strategy~(\ref{P_LHV}). Otherwise, it would be possible to obtain a nonlocal distribution from local operations, which would contradict~(\ref{P_LHV}). Let us divide the set $X$ into two arbitrary subsets named $X_1$ and $X_2$. Then let us use the following notation to represent these subsets:
\begin{align}
X_R &= \{(x_0,0), (x_1,1), (x_2,2)\},\nonumber\\
X_L &= \{(0,x'_0), (1,x'_1), (2,x'_2)\},
\end{align}
where $x_i$ and $x_i'$ can take values in $\{0,1,2\}$. Through the grouping of $X$ into $X_1$ and $X_2$ we see that one of them will contain either $X_L$ or $X_R$ or both of them. It is worthy to note that Bob's nine-input set $Y(y,y') = \{0,1,2\}^2$ has not been partitioned. Consequently, if either subset $X_1$ or $X_2$ contains $X_L$ ($X_R$), a coarse-grained distribution $P(ab|xy)$ on the first copy ($P(a'b'|x'y')$ on the second copy) will certainly correspond to the algebraic maximum of the Magic square game. Since the probability distribution corresponding to $Q(\text{Magic}) = 9$ cannot be obtained with local strategies (where $L(\text{Magic}) = 8$), thus, an upper bound of $L1\text{bit}(\text{Magic}^{\otimes2}) < 81$ is established. Do note that the one-bit classical bound can take only positive integers, resulting in a maximum upper bound of 80. 
\end{proof}

Using more detailed analytical arguments, it is possible to show that $L1\text{bit}(\text{Magic}^{\otimes2})\le 75$. As the bound can be attained with a specific one-bit strategy, the bound is strict, meaning $L1\text{bit}(\text{Magic}^{\otimes2}) = 75$. 

\subsection{One-bit classical bound for the truncated double Magic square game}
\label{sec:magic1bittrunc}

The $\text{Magic}^{\otimes 2}$ inequality consists of 9 inputs and 16 outputs. Let us now remove two inputs from the input set $\{0,1,2\}^2$. Specifically, we remove $\{21,22\}$ on both sides. By doing so, we obtain a Bell inequality with 7 inputs and 16 outputs, which we call $[\text{Magic}^{\otimes 2}]_s$. This game  still remains a pseudo-telepathy game, since the algebraic bound matches the quantum bound, $Q([\text{Magic}^{\otimes 2}]_s) = 49$, and the local bound is $L([\text{Magic}^{\otimes 2}]_s) = 44$ due to an exhaustive search made by enumerating all deterministic strategies. However, $L1\text{bit}([\text{Magic}^{\otimes 2}]_s) = 48$. The proof is analogous to that of the double Magic square game in Sec.~\ref{sec:magic1bit}. We observe that for any partitioning of the 7-element set $X(x,x') = \{00, 01, 02, 10, 11, 12, 20\}$ into two disjoint subsets $X_1$ and $X_2$, one of the subsets will necessarily contain either $X_L$ or $X_R$. The remaining part of the proof is similar to that of double Magic square game. Notice that since the Bell functional $[\text{Magic}^{\otimes 2}]_s$ is symmetric with respect to party exchange, the quantum value of 49 exceeds the bidirectional one-bit classical bound.

\subsection{Fixed directional one-bit classical bound for an asymmetrically truncated double Magic square game}
\label{sec:magic1bittruncasym}

Let us now consider the case, when the communication direction is fixed, and Alice is allowed to communicate one bit to Bob. Once again, we begin with the double Magic square game, but this time, inputs on Alice and Bob respective sides are as follows 
\begin{align}
X(x,x') &= \{00, 01, 02, 10, 11, 12, 20\},\nonumber\\
Y(y,y') &= \{00, 11, 22\}.    
\end{align}
This provides us with a Bell functional having seven inputs on the Alice side and three inputs on Bob's side, with 16 outputs per measurement on both sides. The Bell functional is denoted as $[\text{Magic}^{\otimes 2}]_a$. In this case, the proof follows the same line of reasoning as that for proving the one-bit classical bound $[\text{Magic}^{\otimes 2}]_s$ in Sec.~\ref{sec:magic1bittrunc}. Here, we give different bounds for the expression $[\text{Magic}^{\otimes 2}]_a$, namely, $L1\text{bit}= 20$ (where communication is fixed directional), $Q = 21$ and $L = 18$. Note that a lower bound to the input cardinality in the fixed directional one-bit scenario involves three inputs on Alice's side and two inputs on Bob's side. The question that arises is whether such a Bell-like inequality with quantum violation exists. The answer is affirmative. For this purpose, in Section~\ref{sec:cglmp}, we will recap another family of bipartite Bell inequalities known as the $\text{CGLMP}_d$ family~\cite{CGLMP2002}.  

\subsection{One-bit classical bound for two copies of generic pseudo-telepathy games}
\label{sec:magic1bitpseudo}

In section~\ref{sec:magic1bit}, we observed that by considering a pseudo-telepathy game, specifically the Magic square game, and using two copies of them, we can beat the one-bit classical bound by allowing quantum resources. We ask if it is a generic property of bipartite pseudo-telepathy games. This turns out to be the case. This is because for any bipartition $X_1$ and $X_2$ of Alice's set, $X(x,x') = \{0,...,m_A-1\}^2$, one of the bipartitions will involve $X_L = \{(0,x'_0), ..., (m_A-1,x'_{m_A-1})\}$ or $X_R = \{(x_0,0), ..., (x_{m_A-1},m_A-1)\}$,  or even both, which can be subsequently coarse-grained to the original single-copy pseudo-telepathy game. This property can be shown as follows. 

\begin{proof}
Let us consider partition $X_1$. There are two options. Either it includes $X_L$ or it doesn't. If it does, we have completed the proof. Let us assume that $X_1$ does not include $X_L$. In this case, there is a missing entry $e\not\in\{0,...,m_A-1\}$ on copy $x$. Therefore, $X_2$ needs to contain $e$ on the same copy $x$. Hence, all the pairs $\{(e,0), (e,1), ...,(e,m_A-1)\}$ are included in $X_2$. Consequently, $X_2$ contains $X_R$, where $x_i = e$.
\end{proof}

What is known about two-party pseudo-telepathy games? A number of them have been discussed in the literature~\cite{Brassard2005,Cleve2004}. According to the above proof, the one-bit classical bound can also be violated quantumly with any two-copy pseudo-telepathy game. Let us examine the Impossible coloring game~\cite{Kochen1990,Heywood1983} and for a modern formulation see Ref.~\cite{Cleve2004}. In this game, Alice has three outputs ($o_A = 3$), and Bob has two outputs ($o_B = 2$). The shared state is a $3\times 3$ maximally entangled state. If two copies of the Impossible coloring game are played, the output cardinalities are squared, resulting in $o_A = 9$, $o_B = 4$, and a state space of $9\times 9$. Note that this inequality is not symmetric for party exchange. Since pseudo-telepathy is a property which is symmetric for party exchange, the one-bit classical bound for the double Impossible coloring game is violated quantumly in both directions. Although the dimensionality of the problem (i.e., $d=9$) is less than that of the double Magic square game (which is $d=16$), it requires many more inputs. It is worth noting that this is the smallest output cardinality and dimensionality of the state space that a two-copy bipartite pseudo-telepathy game can have~\cite{Cleve2004}. It should also be noted that there is a correlation Bell inequality with only one bit of communication, which is more economical in terms of dimensionality. In Sec.~\ref{sec:platonic}, we will present the construction for $d = 8$ with 63 inputs and two outputs per party. There is also a construction in the literature based on correlation Bell inequalities for $4\times 4$ systems, which uses an infinite number of inputs~\cite{Vertesi2009}.

\section{Multiple copies of the CGLMP inequality}
\label{sec:cglmp}

In the following subsections, we aim to violate quantumly the one-bit and c-bit bounds with minimum input cardinality using two or more copies of the family of CGLMP inequalities.

\subsection{One-bit bound for the double CGLMP inequalities}
\label{sec:cglmp1bit}

The family of CGLMP inequalities, which was introduced in Ref.~\cite{CGLMP2002}, forms a one-parameter family of bipartite Bell inequalities. Each party has two inputs labelled as 0 and 1, and $d \ge 2$ outputs per party, which are labelled from 0 to $d-1$. Just like in the CHSH and Magic square games, we will use them as basic building blocks of multi-copy inequalities. For $d=2$, they reduce to the CHSH inequality (in an equivalent form) and are tight for all $d$~\cite{Masanes2002}. For $d = 3$, the original three-output CGLMP inequality is recovered~\cite{Acin2002}. We use the form of the inequalities that is presented in Ref.~\cite{Zohren2008}. With a slight modification to the notation in Ref.~\cite{Zohren2008}, we have
\begin{align}
\text{CGLMP}_d =& P(A_0\ge B_0) + P(A_0\le B_1) \nonumber\\
&+ P(A_1 < B_0) + P(A_1 \ge B_1) \le 3,    
\label{cglmpineq}
\end{align}
where $P(A_x < B_y) = \sum_{a<b}P(ab|xy)$, and $L(\text{CGLMP}_d) = 3$ for any $d\ge 2$. Since all coefficients in the inequality~(\ref{cglmpineq}) are positive, this form of $\text{CGLMP}_d$ can be interpreted as a Bell nonlocal game. 

\begin{table}[t]
\begin{center}
\begin{tabular}{ |r|r|r|r| } 
 \hline
$d$ & $Q(\text{CGLMP}_d)$  & $Q(\text{CGLMP}_d^{\otimes 2})$ & $L1\text{bit}(\text{CGLMP}_d^{\otimes 2})$\\
 \hline\hline
2  & 	3.2071  &	10.2855      & 		 12\,\; \\
3  & 	3.3050  &	10.9227      & 		 12\,\; \\
4  & 	3.3648  &	11.3216      & 		 12\,\; \\
5  & 	3.4063  &	11.6028      & 		 12\,\; \\ 
6  & 	3.4374  &	11.8155      &		 12\,\; \\
7  & 	3.4618  &	11.9844      &		 12\,\; \\
8  & 	3.4818  &	12.1230      &		 12\,\; \\
9  & 	3.4985  &	12.2397      &	     12\,\; \\  
10 & 	3.5128  &	12.3399      &		 12\,\; \\      
$\infty$ &  4   &	 	16       &     	 $12^*$\\
 \hline\hline
\end{tabular}
\end{center}
\caption{\label{tab_IV} The table shows the parameter $d$ of the $\text{CGLMP}_d$ inequality, as well as the quantum value $Q$ of the one-copy and two-copy $\text{CGLMP}_d$ inequalities, alongside the one-bit bound $L1\text{bit}$ of the two-copy $\text{CGLMP}_d$, from $d=2$ to $d=10$. The quantum value for $d\to\infty$ is also shown, which is due to Ref.~\cite{Zohren2010}. The entry for $L1\text{bit}$ for $d\to\infty$ is a conjectured value.}
\end{table}

The quantum value $Q(\text{CGLMP}_d)$ up to $d = 10^6$ has been computed by Zohren and Gill~\cite{Zohren2010}, which is believed to be the maximum quantum value, that is, the Tsirelson bound of the inequality for any $d$. The optimal conjectured bipartite quantum state for $\text{CGLMP}_d$ has dimensions $d\times d$. In the second column of Table~\ref{tab_IV}, we reproduce the quantum values up to $d = 10$. For any $d\ge 2$, the algebraic maximum of the single-copy inequality is 4 and the local bound is 3. It has been proven in Ref.~\cite{Zohren2010} that $Q(\text{CGLMP}_d)$ tends to 4 when $d \to\infty$. As a result, in the limit of $d\to\infty$, it provides us with a pseudo-telepathy game. However, as a two-input Bell inequality, the one-bit bound always saturates the algebraic bound, meaning $L1\text{bit}(\text{CGLMP}_d) = 4$. This value is greater than $Q(\text{CGLMP}_d)$ for any finite $d\ge 2$.

Let us now consider playing two or more instances of the $\text{CGLMP}_d$ game in parallel. Table~\ref{tab_IV} (third column) presents the quantum values for the double CGLMP inequalities, where the lower bound $Q(\text{CGLMP}_d^{\otimes 2}) = Q(\text{CGLMP}_d)^2$ is used. When $d = 2$, this formula defines the exact Tsirelson bound since $\text{CGLMP}_2$ is equivalent to CHSH, which is a type of XOR game~\cite{Cleve2008}. The one-bit bound $L1\text{bit}(\text{CGLMP}_d^{\otimes 2}) = 12$ in the last column is verified by the branch-and-bound algorithm up to $d = 10$. We conjecture that this is the exact bound for any $d\ge 2$. We also computed the local bound up to $d = 10$ and obtained $L(\text{CGLMP}_d^{\otimes 2}) = 10$. According to the results in Table~\ref{tab_IV}, $Q(\text{CGLMP}_8^{\otimes 2}) > L1\text{bit}(\text{CGLMP}_8^{\otimes 2})$, therefore we have an example of a four-input, 64-output Bell-like inequality with one bit of communication that can be violated with a $64\times 64$ quantum state.

To achieve $Q(\text{CGLMP}_d)$ in Table~\ref{tab_IV}, non-maximally entangled states of two $d$-dimensional quantum systems are required, except for $d = 2$~\cite{Acin2002}. Let us denote the  quantum value by $\tilde{Q}(\text{CGLMP}_d)$ that can be attained with conjectured optimal measurements and $d\times d$ maximally entangled states~\cite{CGLMP2002,Zohren2008}. The obtained values are $\tilde{Q}(\text{CGLMP}_{31}) = 3.6345$ and $\tilde{Q}(\text{CGLMP}_{31}^{\otimes 2}) = \tilde{Q}(\text{CGLMP}_{31})^2 = 12.0031$. 
Thus, a $(961\times 961)$-dimensional maximally entangled state allows us to exceed the conjectured value of $L1\text{bit}(\text{CGLMP}_{31}^{\otimes 2})=12$. 

At this point, an interesting question arises: What is the minimum number of inputs required to exceed the one-bit bound of a Bell-like inequality with quantum systems? The previous example consists of four inputs. Is there any three-input Bell inequality with one bit of communication violated by quantum systems? We provide such a construction in the next subsection. 

\subsection{One-bit classical bound for truncated double CGLMP inequalities}
\label{sec:cglmp1bittrunc}
Let us consider the double $\text{CGLMP}_d$ inequality of the previous subsection, where we label the four inputs by 
$X(x,x') = \{0,1\}^2$ and $Y(y,y') = \{0,1\}^2$ on the respective sides of Alice and Bob. To obtain the settings, let us remove setting $(1,0)$ from both $X$ and $Y$: 
\begin{align}
X(x,x') &= \{00, 01, 11\}\nonumber\\ 
Y(y,y') &= \{00, 01, 11\}.
\end{align}
Let us denote this three-input inequality by $[\text{CGLMP}_d^{\otimes 2}]_s$. Note that the algebraic maximum is 9 for any $d\ge 2$. This value can be attained quantumly in the limiting case of $d\to\infty$. On the other hand, it can be proven that $L1\text{bit}([\text{CGLMP}_d^{\otimes 2}]_s)$ = 8 for any $d\ge 2$. The proof follows the same line of reasoning as the proof of $L1\text{bit}([\text{Magic}^{\otimes 2}]_s) = 48$ in Sec.~\ref{sec:magic1bittrunc}. In particular, we show that for any bipartition of the three-element set $X(x,x') = \{00, 01, 11\}$, the two-setting $\text{CGLMP}_d$ will appear on one of the partitions. This cannot be played perfectly using local strategies, hence the bound has to be smaller than the algebraic maximum of 9. As all deterministic one-bit strategies produce an integer value, an upper bound for $L1\text{bit}([\text{CGLMP}_d^{\otimes 2}]_s)$ is 8, which is tight since it can attained with a specific one-bit classical strategy. On the other hand, the conjectured local bound is $L([\text{CGLMP}_d^{\otimes 2}]_s) = 7$, which we verified up to $d=20$.

We expect to exceed the bound of $L1\text{bit}([\text{CGLMP}_d^{\otimes 2}]_s)=8$ with a potentially large but finite value of $d$. Why is that? This is due to the fact that $[\text{CGLMP}_d^{\otimes 2}]_s$ can be played perfectly when $d$ is infinite, and its quantum value tends to 9 as $d$ becomes large. For this reason, there must be a threshold value for which $Q([\text{CGLMP}_d^{\otimes 2}]_s)$ exceeds the one-bit bound of 8. Indeed, using the specific settings stated in Ref.~\cite{Zohren2010} and the same quantum states, we obtain $Q([\text{CGLMP}_d^{\otimes 2}]_s) = 8.0002059$ for $d = 283$, exceeding the one-bit bound 8. Hence, we can conclude that a ($283^2\times 283^2$)-dimensional quantum state with well-chosen measurements violates the three-input and $283^2$ output Bell inequality with one bit of communication. 

\subsection{Fixed directional one-bit classical bound for truncated double CGLMP inequalities}
\label{sec:cglmp1bittruncfixed}

We begin with the four-setting $X=Y=\{0,1\}^2$ double CGLMP inequalities by keeping the settings $X(x,x') = \{00, 01, 11\}$ and $Y(y,y') = \{00, 11\}$. We call this expression as $[CGLMP_d^{\otimes 2}]_a$. We allow a single bit of communication from Alice to Bob. Here we find that the one-bit classical bound is 5, and we know that for large $d$ the quantum value converges to 6. The $L1\text{bit}=5$ value is argued similarly to the proof presented in Section~\ref{sec:cglmp1bittrunc}, while the conjectured local bound $L = 4$, which we verified up to $d = 20$.

What is the threshold parameter $d$ at which the quantum value exceeds the one-bit classical bound? It turns out that $Q([CGLMP_d^{\otimes 2}]_a) = 5.0005455617$ for $d = 38$. Therefore, if we consider one bit of classical communication in a fixed direction from Alice to Bob, there is a Bell-like inequality augmented by one bit of communication with three inputs on Alice's side and two inputs on Bob's side that can be violated by a ($38^2\times 38^2$)-dimensional quantum state. 

\subsection{The one-way $c$-bit bound for truncated multicopy CGLMP inequalities}
\label{sec:cglmp1bittrunccbit}

Here we generalize the construction described in section~\ref{sec:cglmp1bittrunc}, from one bit of communication to $c$ bits of communication. We begin with $l=2^c$ copies of the $\text{CGLMP}_d$ expression, and keep the following $(l+1)$ inputs on the respective sides of Alice and Bob: 
\begin{equation}
X = Y = \{0\ldots00, 0\ldots01, 0\ldots11, \ldots, 1\ldots11\}.  
\label{multicglmpXY}
\end{equation}

Let us allow $c$ bits of classical communication from Alice to Bob. With this message, we can make any $l=2^c$-partition of the $(l+1)$-element input set on Alice's side. This will lead to one of the partitions having two inputs. As all $l+1$ strings in (\ref{multicglmpXY}) are different, these will be at least one index, let's say $i$, where these two strings differ. Let us select the same two strings on Bob's side as well. Let us coarse-grain on all the indices, except for index $i$, on both Alice's and Bob's side. This way, we obtain a $\text{CGLMP}_d$ inequality that cannot be played perfectly with only local resources. Therefore, the $c$-bit bound of the truncated $l$-copy $\text{CGLMP}_d$ inequality with $c$ bits of communication from Alice to Bob, is at most $Lc\text{bit} = (l+1)^2-1$, given that the algebraic maximum is $(l+1)^2$. Since the quantum bound of this truncated single copy $\text{CGLMP}_d$ inequality approaches the algebraic maximum when $d\to\infty$, there must be a critical $d$ for any $l=2^c$, where the quantum bound exceeds $Lc\text{bit}$. Note, however, that this critical $d$ can be quite large even for moderate $c$. As we already found in section~\ref{sec:cglmp1bittrunc}, for $c=1$ the critical $d$ is 283. It is worth noting that since the inequality is symmetric for party exchange, our findings regarding exceeding the $c$-bit bound quantumly are still valid when Bob communicates $c$ classical bits to Alice. For a given $c$ number of bits, the number of inputs per party is $2^c+1$ defining a minimal scenario, and both parties have $o = {d^2}^c$ outputs per measurement. Similar results regarding input cardinality have been obtained by Maxwell and Chitambar~\cite{Maxwell2014} concerning the one-way communication cost of simulating no-signalling distributions. In contrast to our case, however, binary outputs could be chosen in that scenario. It remains an open question on how to decrease the dimension and number of outputs by considering alternative Bell inequalities with $c$ bits of communication, where the number of settings is the minimal $2^c+1$.

\section{Platonic correlation-type Bell inequalities with one bit of communication}
\label{sec:platonic}

How do we define a Platonic Bell inequality? Consider an $m$-vertex solid in Euclidean dimension $n$, where the unit vectors pointing towards the vertices of this solid are denoted by $\{\vec V_i\}_i$ with $i=1,...,m$. We suppose that the columns of the matrix, with elements $V_{ij} = (\vec V_i)_j$, are orthogonal to each other and have an equal norm. This property applies to all Platonic and Archimedean solids~\cite{Pal2022}. We shall define the coefficients of the $m$-input two-output correlation Bell inequality 
\begin{equation}
\text{Plato} = \sum_{x=1}^m\sum_{y=1}^m M_{xy}E_{xy} \le L,  
\label{platoineq}
\end{equation}
by $M_{x,y} = \vec V_x\cdot\vec V_y$, where $L$ denotes the local bound, and $E_{xy} = P(00|xy) + P(11|xy) - P(01|xy) - P(10|xy)$ is the two-party correlation between inputs $x$ and $y$. The maximum quantum value of the Bell inequality~(\ref{platoineq}) is $Q(\text{Plato}) = m^2/D$, which defines the Tsirelson bound of the Bell inequality~\cite{BolonekLason2021,Pal2022}. All such Bell inequalities are referred to as Platonic Bell inequalities. One such Platonic construction follows from halving the 126 minimal vectors in the $E_7$ lattice (see the database~\cite{Sloane}). It results in 63 unit vectors $\{V_i\}_i$ in dimension 7, which obey the aforementioned semi-orthogonality property. Hence, the maximum quantum value of this Bell expression is given by $Q(\text{Plato}_{E7}) = m^2/n = 567$. On the other hand, the local bound $L(\text{Plato}_{E7}) = 399$ is obtained through a branch-and-bound search over all local deterministic strategies (see Ref.~\cite{Divianszky2017}. The computation gives the exact local bound. A latest implementation on GPU can be found in~\cite{GPUforL}. The ratio of $Q/L = 567/399 = 1.421052$
exceeds the ratio of the maximum quantum violation of the CHSH inequality, which is $\sqrt{2}$. In contrast, calculating the $L1\text{bit}$ bound is more demanding. For this purpose, we used a heuristic see-saw type iterative algorithm similar to the algorithms described in Ref.~\cite{Araujo2020,Divianszky2017}. This iteratively searches within the set of deterministic one-bit strategies. In this way, we find the best possible lower bound of 563 to $L1\text{bit}(\text{Plato}_{E7})$, which is smaller than $Q(\text{Plato}_{E7})=567$. 

We can actually establish an upper bound of $565$ on $L1\text{bit}(\text{Plato}_{E7})$, which conclusively proves that the quantum value exceeds the one-bit classical bound. Due to Tsirelson's work~\cite{Tsirelson1987} (see also~\cite{Acin2006}), then it is possible to construct 63 projective measurements in dimension $d = 2^{\lfloor{n/2\rfloor}} = 8$ for $n = 7$, along with an $8\times 8$ maximally entangled state state. The analytical bound of $565$ can be shown by using the following observation. 
\begin{observation}\label{obs:L1perL}
The relation 
\begin{equation}
 L1\text{bit}(\text{Plato})\le\sqrt 2 L(\text{Plato})    
\end{equation}
is valid for any Platonic Bell inequality.
\end{observation}
The local and one-bit bounds on the left- and right-hand sides are defined by their respective formulas
\begin{equation}
L(\text{Plato})=\left(\max_{\vec u\in S^{n-1}}\sum_{i=1}^m\abs{\vec u\cdot\vec V_i}\right)^2.    
\label{Lplato}
\end{equation}
and
\begin{align}
&L1\text{bit}(\text{Plato})=\sqrt{L(\text{Plato})}\nonumber\\
&\times\max_{\vec u_1\vec u_2\in S^{n-1}}\sum_{i=1}^m\max\left(\abs{\vec u_1\cdot\vec V_i},\abs{\vec u_2\cdot\vec V_i}\right),    
\label{L1bitplato}
\end{align}
where $\vec V_i\in\R^n$, $i=1,\ldots,m$ are the construction vectors that define the Platonic Bell coefficients as $M_{xy}=\vec V_x\cdot \vec V_y$. One can show the validity of Obs.~\ref{obs:L1perL} by using geometrical arguments based on formulas~(\ref{Lplato},\ref{L1bitplato}). Let us now apply Obs.~\ref{obs:L1perL} to the $\text{Plato}_{E7}$ Bell expression to obtain an upper bound of $L1\text{bit}\le\sqrt{2}\times399 < 565$. As $\text{Plato}_{E7}$ is constructed to be symmetric for party exchange, the one-bit classical bound is the same for both communication directions. Therefore, we conclude that the maximum quantum value of the Platonic Bell inequality $\text{Plato}_{E7}$ above cannot be achieved with a bidirectional one-bit classical communication model. In this case, the dimension of the full probability space is $D_P = 63^2\times 2^2 = 15876$, which is larger than $D_P = 7^2\times 16^2 = 12544$ corresponding to the truncated double Magic square game discussed in Sec.~\ref{sec:magic1bittrunc}.

\section{Discussion}
\label{sec:disc}

We used diverse techniques to prove that a classical model with one bit of classical communication cannot simulate measurements performed on higher dimensional bipartite quantum systems. Table~\ref{tab_I} highlights our main findings for the different constructions. We defined a measure of hardness for one-bit classical simulation by the dimension of the full probability space of the bipartite correlations $D_P$. This is the dimension that is required by quantum correlations to refute classical models with one bit of communication. We placed the upper bound for the value of $D_P$ at 12544. However, this number is quite far from the best lower bound of $D_P > 24$. This suggests that there is still a lot of room for further improvement. We leave it as an open problem to reduce the gap described above. However, it is possible that our attempts to find a smaller upper bound for $D_P$ failed, as its true value might be closer to our upper bound of 12544. If that is the case, we can argue that how surprisingly powerful local hidden variables  models plus a single bit of classical communication are, when the goal is to simulate bipartite quantum correlations. 

From an experimental point of view, it is also crucial to find violations of the one-bit bound using the smallest possible dimensional bipartite states. In this regard, it is known that the double pseudo-telepathy games require at least $9\times 9$ dimensional states. Also, our best construction with a finite number of inputs in terms of dimensionality is based on a Platonic Bell inequality and involves $8\times 8$ dimensional states. On the other hand, there is strong evidence that $2\times 2$ quantum states can be simulated classically with a single bit of communication. The question arises whether one can rule out one-bit classical simulation with a component space dimension less than 8 (and possibly a modest number of inputs) by considering other Bell-like constructions.   

\section*{Acknowledgements}

T.V.~thanks Antonio Ac\'in and Jonatan Bohr Brask for inspiring conversations. We acknowledge the support of the EU (QuantERA eDICT) and the National Research, Development and Innovation Office NKFIH (No. 2019-2.1.7-ERA-NET-2020-00003).

\bibliography{biblios}

\end{document}